\documentclass[12pt,a4paper]{article}
\usepackage{amsmath,amsthm,amsfonts,amssymb,bbm,sansmath}
\usepackage{graphicx,psfrag,subfigure,color}
\usepackage{cite}
\usepackage{hyperref}

\numberwithin{equation}{section}

\newcommand{\Or}{\mathcal{O}}

\newcommand{\Pb}{\mathbbm{P}}

\DeclareMathOperator{\Tr}{Tr}

\newtheorem{prop}{Proposition}[section]

\newtheorem{cla}[prop]{Claim}

\newtheorem{rem}[prop]{Remark}

\title{Coupled Kardar-Parisi-Zhang Equations\newline in One Dimension}
\author{Patrik L.\ Ferrari\footnote{Institute for Applied Mathematics, Bonn University, Endenicher Allee 60, 53115 Bonn, Germany. E-mail: {\tt ferrari@uni-bonn.de}} \and
Tomohiro Sasamoto\footnote{Mathematics Department, Chiba University, Yayoi--cho 1--33, Inage, Chiba 263--8522, Japan. E-mail: {\tt sasamoto@math.s.chiba-u.ac.jp}}$\,\,^\ddagger$
\and
Herbert Spohn\footnote{Zentrum Mathematik, TU M\"unchen, Boltzmannstrasse 3, D-85747 Garching, Germany. E-mails: {\tt spohn@ma.tum.de} and {\tt sasamoto@ma.tum.de}}}

\date{22. August 2013}

\begin{document}
\maketitle
\sloppy
\vfill
\begin{abstract}
Over the past years our understanding of the scaling properties of the solutions to the one-dimensional KPZ equation has advanced considerably, both theoretically and experimentally. In our contribution we export these insights to the case of coupled KPZ equations in one dimension. We establish equivalence with nonlinear fluctuating hydrodynamics for multi-component driven stochastic lattice gases. To check the predictions of the theory, we perform Monte Carlo simulations of the two-component AHR model. Its steady state is computed using  the matrix product ansatz. Thereby all coefficients appearing in the coupled KPZ equations are deduced from the microscopic model.
Time correlations in  the steady state are simulated and we confirm not only the scaling exponent, but also the scaling function and the non-universal coefficients.
\end{abstract}

\vfill

\section{Introduction}\label{SectIntro}
In the early 90ies Erta\c{s} and Kardar~\cite{EK92,EK93,Kar98} studied the dynamic roughening of directed lines, as for example dislocation, polymer, or vortex lines, and for that purpose used a model  consisting of two coupled one-dimensional KPZ equations. Prominent further examples, employing the same type of Langevin equations, are sedimenting colloidal suspensions~\cite{LRFB98} and crystals~\cite{RS97}, stochastic lattice gases~\cite{DBBR01}, and magnetohydrodynamics~\cite{FD98,Yan97,BBR99}.  The Langevin equation has the generic structure
\begin{equation}\label{1a}
\partial_t h_\alpha = -v_\alpha  \partial_x h_\alpha + \sum_{\beta,\gamma= 1}^n G^\alpha_{\beta\gamma}( \partial_x h_\beta )(
 \partial_x h_\gamma) +  \sum_{\beta= 1}^n  D_{\alpha\beta} \partial_x^2  h_\beta  +
  \sum_{\beta= 1}^n  B_{\alpha\beta}\xi_\beta.
\end{equation}
$\alpha$ labels the fields, $\alpha = 1,\ldots,n$. The components, $h_\alpha(x,t)$, have varying interpretations depending on the physical context
and we refer to them simply as $\alpha$ component of the height. $v_\alpha$ is the imposed drift velocity of the
$\alpha$th component.  $G^\alpha_{\beta\gamma}$ is the coupling of the  time change of height $\alpha$ to the slope of the heights $\beta$ and $\gamma$. Obviously $G^\alpha_{\beta\gamma}= G^\alpha_{\gamma\beta}$. $D$ is the diffusion matrix and $BB^\mathrm{T}$ is the noise strength matrix, where $\xi_\alpha(x,t)$ are independent, space-time white noises with covariance
\begin{equation}\label{1b}
\langle \xi_\alpha(x,t) \xi_{\alpha'}(x',t')\rangle =  \delta_{\alpha\alpha'}  \delta(x - x')  \delta(t - t').
\end{equation}
For $n=1$ the index $1$ is omitted and $h_1 = h$.  Then (\ref{1a}) is the original KPZ equation~\cite{KPZ86}.
Note that the velocity $v_1$ can be removed by switching to a co-moving reference frame. But for $n > 1$
this is no longer possible, in general.

In~\cite{EK92} the following particular case was considered: $n=2$, $v_\alpha = 0$, $D$, $B$ diagonal, and $G^1_{12} = 0$,
 $G^2_{11}
= 0 = G^2_{22}$.
In further studies,  also three components, non-zero velocities, and further couplings are considered. Of particular interest for us is the link to multi-component driven lattice gases. This context will be explained in fair detail below, but to mention already now, for lattice gases $\partial_x h_\alpha$ will be the $\alpha$th normal mode and  the velocities $v_\alpha$ differ from each other, generically. We refer to~\cite{DBBR01} for a most illuminating case study.

Much  more recently~\cite{vanB12,MS13,Spo13}, it has been realized that the coupled KPZ equations (\ref{1a}) cover also the dynamics of
fluids in one dimension and of anharmonic chains on a mesoscopic scale. These systems have three conserved fields, number, momentum, and energy, hence $n=3$. Langevin equations of the form (\ref{1a}) govern the coupled dynamics
consisting of two sound modes and the heat mode, labeled canonically by $\alpha = -1,0,1$. One finds $v_0 =0$,
$v_{\pm1} = \pm c$ with $c$ the speed of sound.
In particular, $G^0_{00} = 0$ always, which implies that the heat mode will scale differently from the sound modes, see~\cite{vanB12,Spo13} for details. For anharmonic chains one can add stochastic collisions either respecting all or only some of conservation laws~\cite{BO11}. Such stochastic models on a mesoscopic scale are also described by Eq.~(\ref{1a}), see e.g.~\cite{BS12,BeGo13}.

Of central interest are the steady state time correlations $\langle \partial_x h_\alpha(x,t)\partial_x h_\beta(0,0)\rangle$
for (\ref{1a}). Depending on the physical application other correlations and/or initial conditions could be considered.
One obvious example are deterministic initial conditions as the flat data $h_\alpha(x,0) = 0$. Mostly in the context of one-dimensional fluids,  one studies (\ref{1a}) restricted to a finite interval, $x \in [0,L]$, with dynamically prescribed boundary fluxes at both ends. Then the steady state profile and its fluxes would be the prime target. Such applications are beyond the present scope and we concentrate on the steady state time correlations.

The case of a single component is very well understood. For a detailed summary we refer to Section~\ref{single}.
For $n > 1$,
the pioneering works quoted above focus on the properties of the steady state for (\ref{1a}) and on the scaling exponents  of the heights. Now larger systems can be studied and finer details are resolved. In particular, molecular dynamics for anharmonic chains has been carried out by  several groups with great intensity~\cite{Lep03,Dha08,CZWZ13}. Following the example of a single component, one would like to identify the universality classes, to predict and to numerically compute the universal scaling exponents and functions, and to express the nonuniversal coefficients in terms of $v_\alpha$, $G^\alpha$. This is a grand program and our contribution is only a very first step. We will discuss the relation between driven stochastic lattice gases and coupled KPZ equations. As benchmark example we
perform Monte Carlo simulations of  the two-component AHR model~\cite{AHR98,AHR99}, for which $n = 2$ and $v_1 \neq v_2$. The steady states  have to be computed through the matrix product ansatz and are generically not of product form, thereby providing a more severe test of the theory.

In Section~\ref{SectHydro} we first recall the case of a single conserved field and  then introduce a $n$-component lattice gas on $\mathbb{Z}$. Its fluctuating hydrodynamics will be transformed into $n$ coupled KPZ equations. In Section~\ref{SectAHR} we explain the matrix product ansatz for the AHR model.
In particular we obtain analytically $v_\alpha$ and the matrices $G^\alpha$ in their dependence on the densities of the two components.
The respective
Monte Carlo simulations are reported in Section~\ref{SectMCSimulation}.
It turns out that for the AHR model the theoretically predicted scaling is established already on fairly short time scales.
This is a rather surprising observation in view of the experience with anharmonic chains, for which slow convergence seems to be  the rule.  As obvious difference, the AHR model is governed by a stochastic dynamics, while anharmonic chains are deterministic following the hamiltonian equations of motion. Since the mesoscopic nonlinear fluctuating hydrodynamics is identical in structure, one has to search for a deeper explanation, which will be discussed in Section~\ref{SectCorrectionScaling}.

\section{Multi-component systems and fluctuating hydrodynamics}\label{SectHydro}
\subsection{A single mode}\label{single}
 Let us first recall the very well understood case of a single component, \mbox{$n=1$}, since it will serve as a blueprint for the general case.  In fact, we will concentrate on the simplest model, namely the totally asymmetric simple exclusion process (TASEP) which consists of  particles hopping on the one-dimensional lattice $\mathbb{Z}$. At each site there is at most one particle, the occupation variable  of site $j$ being denoted by $\eta(j)$, $\eta(j) \in \{0,1\}$. Particles hop independently to the right with rate 1. Jumps violating the exclusion rule are suppressed. The steady state of the TASEP is characterized by an average density $\rho$, $0 \leq \rho \leq 1$. In the steady state the $\eta(j)$'s are independent and $\mathbb{P}(\eta(j) = 1) =\rho$. Running the stochastic dynamics with this initial condition defines the space-time stationary process $\eta(j,t)$
and our interest are the correlations of the conserved field,
\begin{equation}\label{1}
S(j,t) = \langle \eta(j,t)\eta(0,0)\rangle_\rho - \rho^2,
\end{equation}
where $\langle \cdot \rangle_\rho$ refers to the average in the stationary process with density $\rho$. There are two general identities which follow directly from the conservation law. Defining the susceptibility
\begin{equation}\label{2a}
\chi(\rho)= \sum_{j \in \mathbb{Z}} S(j,0),
\end{equation}
$\chi=\chi(\rho) = \rho(1-\rho)$ for TASEP, it holds
\begin{equation}\label{2}
\sum_{j \in \mathbb{Z}} S(j,t) = \chi.
\end{equation}
In addition, see \cite{PS01} e.g.,
\begin{equation}\label{3}
\chi^{-1}\sum_{j \in \mathbb{Z}} jS(j,t) = \mathsfsl{j}'(\rho)t,
\end{equation}
where $\mathsfsl{j}(\rho)$ is the average current in the steady state, $\mathsfsl{j}(\rho) = \rho(1-\rho)$ for TASEP.
Thus $S(j,t)$ has total weight $\chi$ and is centered at $\mathsfsl{j}'(\rho)t $.

\begin{figure}
\begin{center}
\includegraphics[height=4cm]{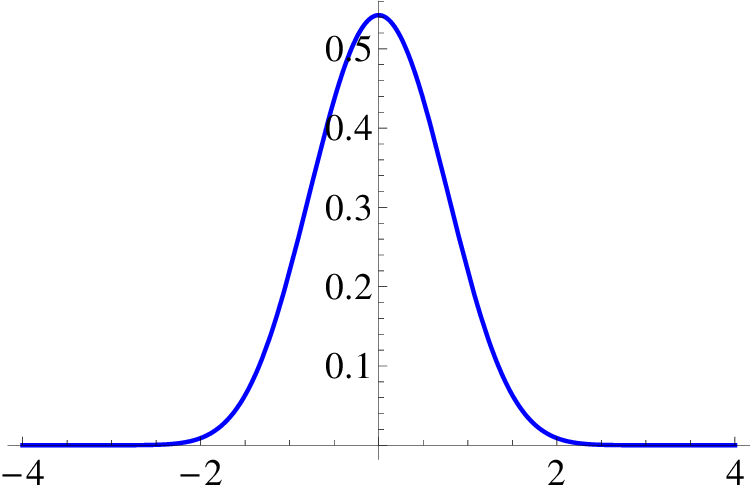}
\caption{Plot of the scaling function $f_{\rm KPZ}$, tabulated in~\cite{PSKPZ}, with the properties
\mbox{$f_{\mathrm{KPZ}} >0$}, $f_{\mathrm{KPZ}}(x) = f_{\mathrm{KPZ}}(-x)$, $\int dx f_{\mathrm{KPZ}}(x) =1$,
$\int dx f_{\mathrm{KPZ}}(x) x^2 =0.51$, and $f_{\mathrm{KPZ}}(x) \simeq e^{-0.3|x|^3}$ for
$|x| \to \infty$.}
\label{FigfKPZ}
\end{center}
\end{figure}

It has been noted already some time ago, that the broadening of $S$ is proportional to $t^{2/3}$, hence superdiffusive
\cite{BKS85,KPZ86}. This is special for one dimension. In higher dimensions one would find diffusive spreading~\cite{EMY94} with $d=2$ as upper critical dimension. In fact, the TASEP allows for an exact solution and in~\cite{PS01,FS05a,BFP12} it is proved that $S$ scales as
\begin{equation}\label{4}
 S(j,t) = \chi (\lambda_\mathrm{0} t)^{-2/3} f_{\mathrm{KPZ}}\left((\lambda_\mathrm{0} t)^{-2/3}(j -  \mathsfsl{j}'(\rho)t)\right)
\end{equation}
for large $j,t$. The non-universal coefficient is $\lambda_\mathrm{0} = \sqrt{2 \chi} |\mathsfsl{j}''(\rho)|$.
$f_{\mathrm{KPZ}}$ is a universal scaling function, see Figure~\ref{FigfKPZ}.

While (\ref{4}) is proved for the TASEP and a few other models, it should hold at much greater generality, where $\chi$ is still defined as in (\ref{2}) and $\mathsfsl{j}(\rho)$ is the steady state current at density~$\rho$.

One argument supporting universality relies on the observation that on the mesoscopic scale, the fluctuations in the density are governed by a stochastic field theory. To derive it one starts from the macroscopic conservation law as
\begin{equation}\label{5}
\partial_t\tilde{u} + \partial_x \mathsfsl{j}(\tilde{u}) = 0
\end{equation}
with $\tilde{u}(x,t)$ the macroscopic density profile. For the standard Gaussian fluctuation theory one would expand at
$\rho$ as $\tilde{u} = \rho + u$ to linear order in $u$ and add phenomenologically dissipation and noise as
\begin{equation}\label{6}
\partial_t u + \partial_x \big( \mathsfsl{j}'(\rho) u - D \partial_x u + \sqrt{2 D\chi} \xi \big) = 0,
\end{equation}
where $\xi$ is normalized space-time white noise. Note that the invariant measure for (\ref{6}) is spatial white noise
with variance $\chi$. If the steady state of the microscopic model has rapidly decaying correlations, this would precisely
describe  its statistical properties on a large spatial scale. In this sense the static properties of (\ref{6}) are consistent with the microscopic model. On the other hand, (\ref{6}) predicts a Gaussian peak traveling with velocity $\mathsfsl{j}'(\rho)$ and of width
$\sqrt t$, in contradiction to (\ref{4}). The crucial insight is that the superdiffusive spreading results from the second order expansion of the current in (\ref{6}), to say the correct, now nonlinear, fluctuating hydrodynamics reads
\begin{equation}\label{7}
\partial_t u + \partial_x \big( \mathsfsl{j}'(\rho) u + \tfrac{1}{2}\mathsfsl{j}''(\rho) u^2- D \partial_x u + \sqrt{2 D\chi} \xi \big) = 0,
\end{equation}
usually called noisy Burgers equation.
Despite adding the nonlinearity $u^2$, spatial white noise of variance $\chi$ is still the stationary measure for (\ref{7})~\cite{KPZ86,FQ13}.  Furthermore the covariance of the stationary mean zero process, $\langle u(x,t)u(0,0)\rangle$, scales as stated in  (\ref{4}). This can be checked through the replica method~\cite{IS12, IS13}. The validity of the scaling (\ref{4}), or related type of scaling behavior, has been confirmed in experiments~\cite{TS10,TS12,MMMT05} and in numerical simulations~\cite{Tak12,AOF11,HH12}.

Integrating in $x$, $\partial_x h = u$, and switching to the frame moving with velocity $\mathsfsl{j}'(\rho)$,  Eq.~(\ref{7}) turns into
\begin{equation}\label{7a}
\partial_t h = -   \tfrac{1}{2}\mathsfsl{j}''(\rho)( \partial_x h)^2 + D \partial_x^2 h + \sqrt{2 D\chi} \xi,
\end{equation}
which is the KPZ equation, \textit{i.e.} Eq.~(\ref{1a}) for $n=1$.

From a mathematical perspective, the KPZ equation (\ref{7a}) is fairly singular. This can be understood by considering the Gaussian process obtained by setting $\mathsfsl{j}''(\rho) = 0$. Then, with probability one, the spatial dependence
$ x \mapsto h(x,t)$ is H\"{o}lder $1/2$, certainly not differentiable. The nonlinearity requires to multiply
pointwise the slope, which is not a well-defined operation. In a recent contribution~\cite{Hai11} Hairer uses the theory of rough paths to give a meaning to the one-component KPZ equation. The case of several components has still to be studied. A related issue is to derive the KPZ equation from a stochastic particle model in a suitable limit. A generic method consists in the limit of weak drive. This is easily illustrated for the PASEP (partially asymmetric simple exclusion process), where particles hop with rate $p$ to the right and rate $q= 1-p$ to the left, still satisfying the exclusion rule. Weak asymmetry means $p = \tfrac{1}{2} + \sqrt{\epsilon}$,
$ 0 < \epsilon \ll 1$. The PASEP height is implicitly defined through $h^\epsilon (j+1,t) - h^\epsilon (j,t) =\eta(j,t)$.
One considers $j =  \mathcal{O}(\epsilon^{-1})$ and $t =  \mathcal{O}(\epsilon^{-2})$. Then the height is $ \mathcal{O}(\epsilon^{-1/2})$. The correspondingly scaled height converges in the limit $\epsilon \to 0$ to the solution of the KPZ equation~\cite{BG97,SS10b,ACQ10}. A corresponding theorem is missing for the multi-component case.

The proofs cited above rely on the integrable structure of the underlying stochastic model, like determinantal point processes, Bethe ansatz, Schur processes, and their generalization on the level of Macdonald processes~\cite{BC11}.
The extension of such techniques to several components has been explored only for a few models~\cite{Weh10,AKSS09,ADHR93,PFS02}.

\subsection{$n$-component lattice gas}\label{sublattice}

The scheme to be described is fairly general. But to  keep track of the assumptions, it is convenient to
consider a concrete class of stochastic lattice gases. The AHR model will then be a particular example.

We consider a lattice gas having $n$ components with particles hopping on the one-dimensional lattice $\mathbb{Z}$.
The components are labeled by $\alpha = 1,\ldots,n$. The position of the $\ell$-th particle of type $\alpha$ at time $t$ is denoted by $x_{\ell,\alpha}(t)$. The particles hop according to a jump process with rates which are not written out explicitly here.
We allow only for nearest neighbor jumps, \textit{i.e.}, hops from the current location  $x_{\ell,\alpha}(t)$ to
$x_{\ell,\alpha}(t)\pm 1$. The jump rates are local and invariant under translations. They are independent of the
particle label $\ell$, but will depend in general on the type $\alpha$. By construction, particles of type $\alpha$ are
conserved. To define the steady states one starts from a ring of $L$ sites with periodic boundary conditions and fixes
the number $N_\alpha$ of $\alpha$ particles. It is assumed that the dynamics explores the whole set of configurations at given $\vec{N}$.
There is then a unique invariant measure $\mu_{\vec{N}}$ and it is also ensured that there are no further conservation laws. We now take the infinite volume limit of $\mu_{\vec{N}}$ through $L, N_\alpha\to \infty$ at fixed $\vec{\rho} =  \vec{N}/L$. In principle the system could phase segregate. Formally this will be excluded by assuming that $\mu_{\vec{\rho}}$, the infinite volume steady state at density $\vec{\rho}$, is translation invariant, ergodic with respect to translation, and has truncated correlations which decay to 0 as the separation of the reference points goes to infinity. Such properties can be  handled only case by case with no general abstract argument available, in sharp contrast to lattice gases satisfying detailed balance~\cite{Geo79}.
In the following we fix $\vec{\rho}$ and assume the above properties for  $\mu_{\vec{\rho}}$ including a small neighborhood of $\vec{\rho}$. The object of interest is the stationary process, denoted again by $x_{\ell,\alpha}(t)$, with initial measure $\mu_{\vec{\rho}}$.
In our notation the dependence on the reference density $\vec{\rho}$ will be mostly suppressed.

Next we introduce the conserved fields
\begin{equation}\label{8}
\eta_\alpha(j,t) = \sum_\ell \delta(x_{\ell,\alpha}(t) - j),
\end{equation}
which is the number of particles of type $\alpha$ at $(j,t)$. (For the TASEP one identifies $\eta_1(j,t)$
with $\eta(j,t)$ from Section~\ref{single}.) $\eta_\alpha$ is a stationary process and by assumption
\begin{equation}\label{9}
\lim_{m \to\infty} \frac{1}{2m} \sum_{j = -m +1}^m \eta_\alpha(j,t) = \rho_\alpha
\end{equation}
almost surely. We also introduce the $\alpha$-th instantaneous current $\mathcal{J}_\alpha(j,t)$. As a function of $t$ it consists of
$\delta$-spikes with weight $1$ located at time instances when a particle of type $\alpha$ jumps from $j$ to $j+1$ and with weight $-1$ at time instances when a particle of type $\alpha$ jumps from $j+1$ to $j$. Mass conservation is then expressed through
\begin{equation}\label{10}
\frac{d}{dt} \eta_\alpha(j,t)  - \mathcal{J}_\alpha(j-1,t) + \mathcal{J}_\alpha(j,t) = 0.
\end{equation}
In the steady state
\begin{equation}\label{11}
\langle  \mathcal{J}_\alpha(j,t) \rangle_{\vec{\rho}} = \mathsfsl{j}_\alpha(\vec{\rho}),
\end{equation}
which is the $\alpha$-th average current. $\vec{\mathsfsl{j}}$ is assumed to depend smoothly on $\vec{\rho}$ in a small neighborhood of the reference density.

The time correlations of interest are codified by
\begin{equation}\label{12}
S_{\alpha\beta}(j,t) = \langle  \eta_\alpha(j,t) \eta_\beta(0,0)\rangle_{\vec{\rho}}  - \rho_\alpha\rho_\beta.
\end{equation}
It will be convenient to regard $S(j,t)$ as a $n \times n$ matrix. Then, as for $n = 1$,
\begin{equation}\label{13}
\sum_{j \in \mathbb{Z}} S(j,t) = \sum_{j \in \mathbb{Z}} S(j,0) = C,
\end{equation}
which defines the susceptibility matrix $C$. Clearly, $C$ is symmetric and non-negative. To avoid a completely frozen component, we require that $C$ has strictly positive eigenvalues, \textit{i.e.}, $C > 0$. As discussed in the Appendix \ref{AppB}, using
space-time stationarity, for the first moment one has the identity
\begin{equation}\label{14}
\sum_{j \in \mathbb{Z}} jS(j,t) =  \sum_{j \in \mathbb{Z}} jS(j,0) + ACt,
\end{equation}
where
\begin{equation}\label{15}
A_{\alpha\beta} (\vec{\rho})= \frac{\partial}{\partial\rho_\beta}  \mathsfsl{j}_\alpha (\vec{\rho}).
\end{equation}
$A$ is regarded as a $n\times n$ matrix.
As a consequence
\begin{equation}\label{15a}
AC = CA^{\mathrm{T}}
\end{equation}
with $^\mathrm{T}$ denoting transpose. Eq.~(\ref{15a}) ensures  in particular that $A$ has real eigenvalues.
The dependence on the background density $\vec{\rho}$ will be mostly suppressed from our notation.

$S_{\alpha\beta}(j,t)$ can be viewed as the average density of $ \eta_\alpha(j,t)$ caused by an initial
perturbation of the $\beta$ density at the origin. On large scales one might hope to capture such a response by a hydrodynamic theory. The first step is to consider density fields  $\tilde{u}_\alpha(x,t)$ varying on the macroscopic scale.
They satisfy the system of conservation laws
\begin{equation}\label{16}
\partial_t \tilde{u}_\alpha(x,t) +\partial_x \mathsfsl{j}_\alpha(\vec{\tilde{u}})(x,t) = 0.
\end{equation}
Linearizing as $\vec{\tilde{u}} = \vec{\rho} + \vec{u}$ one arrives at
\begin{equation}\label{17}
\partial_t u_\alpha + \partial_x (A\vec{u})_\alpha = 0
\end{equation}
with $A$ of (\ref{15}). On this level of precision
\begin{equation}\label{18}
S(j,t)\simeq (e^{At}C)(j,t),
\end{equation}
which implies, denoting by $\{v_\alpha, \alpha = 1,\ldots,n\}$ the eigenvalues of $A$, that $S_{\alpha\beta} (j,t)$
has $n$ peaks centered at $v_\alpha t$. Of course, some of the $v_\alpha$'s might coincide and some of the peaks might be missing because of particular symmetries. With this background information the real issue becomes to understand the broadening of the peaks because of dissipation and noise.

The Gaussian fluctuation theory (\ref{6}) easily extends to $n$ components as
\begin{equation}\label{20a}
\partial_t u_\alpha + \partial_x \big( (A \vec{u})_\alpha
- \partial_x (\tilde{D}\vec{u})_\alpha + (\tilde B \vec{\xi}\,)_\alpha \big) = 0.
\end{equation}
Here $\tilde B\tilde B^\mathrm{T}$ is the noise strength matrix and $\tilde{D} =  \tilde{D}^\mathrm{T}$ is the diffusion matrix.
$\tilde B$, $\tilde{D}$ are constrained by
\begin{equation}\label{22}
\tilde{D}C + C\tilde{D} = \tilde B\tilde B^\mathrm{T}.
\end{equation}
The Gaussian fluctuation theory fails already for a single component.  Hence, as for $n=1$, we expand the Euler currents up to second order yielding the Hessians
\begin{equation}\label{19}
H^\alpha_{\beta\gamma}(\vec{\rho}) = \frac{\partial^2}{\partial\rho_\beta \partial \rho_\gamma}
 \mathsfsl{j}_\alpha(\vec{\rho}).
\end{equation}
Adding to (\ref{20a}) one arrives at the nonlinear fluctuating hydrodynamics
\begin{equation}\label{20}
\partial_t u_\alpha + \partial_x \big( (A \vec{u})_\alpha + \tfrac{1}{2}\langle \vec{u},H^\alpha\vec{u}\rangle
- \partial_x (\tilde{D}\vec{u})_\alpha + (\tilde B \vec{\xi}\,)_\alpha \big) = 0,
\end{equation}
where $\langle\cdot,\cdot\rangle$ denotes the inner product in component space. Since $ \eta_\alpha(j,t)$  is stationary, we will study the space-time stationary solution  to (\ref{20}), here denoted by
$\vec{u}(x,t)$, with $\langle \vec{u}(x,t) \rangle = 0$ since a small deviation from uniformity is considered.

Using $C = C^\mathrm{T} >0$ and the identity (\ref{15a}), $A$ has  real eigenvalues and a non-degenerate system of left and right eigenvectors.
Therefore one can introduce the normal mode coordinates $\phi$ through
\begin{equation}\label{24}
\vec{\phi} = R \vec{u} ,
\end{equation}
such that
\begin{equation}\label{25}
RAR^{-1} = \mathrm{diag}(v_1,\ldots,v_n).
\end{equation}
In addition we require the normal modes to be orthonormal in the steady state, which means
\begin{equation}\label{25a}
RCR^\mathrm{T} = 1.
\end{equation}
Up to an overall factor of $-1$, the transformation matrix $R$ is then uniquely defined.
The $R$ matrix acts only in component space. Therefore the definition (\ref{24}) applies
also to the Fourier transformed fields and to the currents. In the same way, one introduces
the normal mode representation of the microscopic fields.

In normal modes Eq.~(\ref{20}) reads
\begin{equation}\label{26}
\partial_t \phi_\alpha + \partial_x \big( v_\alpha \phi_\alpha + \langle \vec{\phi},G^\alpha\vec{\phi}\rangle
- \partial_x (D\vec{\phi})_\alpha + (B \vec{\xi}\,)_\alpha \big) = 0,
\end{equation}
where $D = R\tilde{D}R^{-1}$, $B=R \tilde B$, and the coupling constants, $G^\alpha$, of the quadratic nonlinearity read
\begin{equation}\label{27}
G^\alpha = \sum_{\alpha' = 1}^n \tfrac{1}{2} R_{\alpha\alpha'} (R^{-1})^\mathrm{T}H^{\alpha'}R^{-1}.
\end{equation}
If one sets $\phi_\alpha = \partial_x h_\alpha$, then Eq.~(\ref{26})  is identical to the coupled KPZ equations (\ref{1a}). Only for the noise strength we have the relation $BB^{\mathrm{T}} = 2 D$, because of $\tilde{D} = \tilde{D}^\mathrm{T}$,
(\ref{22}), and  the orthonormality condition (\ref{25a}).

In contrast to the case $n=1$, we can no longer rely on exact solutions. On the other hand, if
the mode $\alpha$ has its velocity $v_\alpha$  distinct from all other modes, then their separation grows linearly in $t$.
In Fourier space the couplings to the other modes contain rapidly oscillating factors, which
for long times and small wave vectors cancel out the interaction. Hence the dynamics decouples and $\phi_\alpha$
satisfies in approximation
\begin{equation}\label{27a}
\partial_t \phi_\alpha + \partial_x \big( v_\alpha \phi_\alpha + G^\alpha_{\alpha\alpha}\phi_\alpha^2
- \partial_x D_\alpha\phi_\alpha + \sqrt{2D_\alpha}\,\xi_\alpha \big) = 0.
\end{equation}
Following the discussion above one first has to transform to normal modes as
\begin{equation}\label{28}
S_{\alpha\beta}^{\sharp}(j,t) = (RSR^\mathrm{T}) _{\alpha\beta}(j,t) = \langle  (R\vec{\eta})_\alpha(j,t)(R\vec {\eta})_\beta(0,0)\rangle_{\vec{\rho}}  -(R\vec{ \rho})_\alpha (R\vec{\rho})_\beta.
\end{equation}
This matrix should be approximately diagonal and, for large $x,t$,
\begin{equation}\label{28a}
S_{\alpha\alpha}^{\sharp}(j,t)\simeq \langle\phi_\alpha(x,t)\phi_\alpha(0,0)\rangle\\
 \simeq (\lambda_\alpha t)^{-2/3} f_{\mathrm{KPZ}}\left((\lambda_\alpha t)^{-2/3}(x -  v_\alpha t)\right)
\end{equation}
where $f_{\mathrm{KPZ}}$ is plotted in Fig. \ref{FigfKPZ} and $\lambda_\alpha = 2 \sqrt{2} |G^\alpha_{\alpha\alpha}|$.  Compared to Eq.~(\ref{4}) the factor $\chi$ is missing, since by construction $\int dx  \langle\phi_\alpha(x,t)\phi_\alpha(0,0)\rangle = 1$.
If $G^\alpha_{\alpha\alpha} \neq 0$ and if the $v_\alpha$'s are distinct, then the $\alpha$-th normal mode is expected to be governed by KPZ universality.

Eq.~(\ref{28}) is a checkable prediction for stochastic lattice gases. One needs the average currents $\vec{\mathsfsl{j}}$ and the susceptibility matrix $C$. Then one performs the linear transformation $(R\vec{\eta})_\alpha$ to obtain the $\alpha$-th normal mode.
As an important null test, it should be uncorrelated with the other modes and more specifically it should scale as stated in Eq.~(\ref{28a}). As to be emphasized not only the power law $t^{2/3}$ is claimed. Predicted are also the full scaling function and the non-universal coefficient $\lambda_\alpha$.

In the following two sections we explain how the theory is applied to the AHR model. Two steps are required to have a check on the scaling (\ref{28}). Firstly one has to compute $A,R,C$ and $G$, which is done in Section~\ref{SectAHR}. Secondly one has to run Monte Carlo simulations of the AHR model, which is reported in Section~\ref{SectMCSimulation}.

\section{The AHR model}\label{SectAHR}
As an example to apply our general theory from the previous section,
we consider a two-species asymmetric simple exclusion process introduced
by Arndt, Heinzel and Rittenberg~\cite{AHR98,AHR99}. In this AHR model, each site is either occupied by a
$+$ particle, a $-$ particle or is empty (denoted by 0). The hopping and exchange rates are summarized as
\begin{equation}
\begin{aligned}
+~ 0 &\stackrel{ \beta }{\rightarrow} 0~ + \\
0~ - &\stackrel{ \alpha }{\rightarrow} -~ 0  \\
+~ - &\stackrel{1}{\underset{ q }{ \rightleftharpoons }} -~ +
\end{aligned}
\end{equation}
Here we assume $\alpha,\beta>0$ and $0\leq q <1$ with all other rates to be set equal to zero.
In words, each $+$ (resp.~$-$) particle have a tendency to hop to the right (resp.~left) with hopping rate
$\beta$ (resp.~$\alpha$). But if the $+$ and $-$ particles are on neighboring sites, the exchange
happens with rate 1 when the exchange is in the direction compatible with the hopping above
and with rate $q$ for the reverse exchange. It is obvious from the definition that the process
has two conserved quantities, namely the number of $+$ and $-$ particles.
For a comparison with the discussions in the previous section, the AHR model is a $n=2$ component
system with labels $1,2$ renamed as $+,-$. Note that the exclusion rule is encoded in the hopping and
jump rates of the model.

The stationary measure of this process has a special feature as it can be constructed
explicitly by using the matrix product method~\cite{DEHP93,AHR99,RSS00}.
This construction works for the process with periodic boundary condition with system size $L$.
For this setting a configuration $\eta$ of the system can be specified by two sets of binary sequences,
$\eta_{\pm} = \{ \eta_{\pm}(1),\ldots ,\eta_{\pm}(L)\}$
where \mbox{$\eta_{\pm}(j)\in\{0,1\}$} denotes the number of $\pm$ particles at site $j$,  $1\leq j\leq L$.
Note that because of exclusion rule $\eta_+(j) +\eta_- (j)
\in \{0,1\}$ and  we set \mbox{$\eta_0(j)=1-\eta_+(j)-\eta_-(j)$}.
Hence, let us denote the number of
particles of $+$ and $-$ particles by $N_+$ and $N_-$ respectively. When $N_++N_-=L$,
there is no empty site for which case the process becomes actually the ASEP and its
stationary state is known to be product. Let us consider only the case in which
$N_+ + N_-<L$. In this case, the probability $\Pb_\mathrm{can}(\eta)$ that the system is in the configuration
$\eta$ in the stationary state is given by
\begin{equation}
 \Pb_\mathrm{can}(\eta) = \frac{1}{Z_{L,N_+,N_-}} \Tr \prod_{j=1}^L
               \left(  \eta_+(j)\mathsfsl{D}+\eta_-(j)  \mathsfsl{E} + \eta_0(j)  \mathsfsl{A}  \right)
\label{MPA_C}
\end{equation}
where $\mathsfsl{D},\mathsfsl{E},\mathsfsl{A}$ has to satisfy
\begin{equation}
\label{DEA}
\mathsfsl{D}\mathsfsl{E} - q \mathsfsl{E}\mathsfsl{D} = \mathsfsl{D}+\mathsfsl{E}, \quad
\beta \mathsfsl{D}\mathsfsl{A} = \mathsfsl{A},\quad \alpha \mathsfsl{A} \mathsfsl{E} = \mathsfsl{A}
\end{equation}
and $Z_{L,N_+,N_-}$ is the normalization constant. This is a sort of canonical ensemble
because the particle numbers are fixed.

In the following discussions, however, it turns out to be useful to switch to a grand canonical ensemble
with fugacities $\xi_+, \xi_-$ for which the probability $\Pb(\eta)$ is given by
\begin{equation}\label{eq3.4}
 \Pb(\eta) = \frac{1}{Z_L(\xi_+,\xi_-)}
 \Tr \prod_{j=1}^L \left(  \eta_+(j) \xi_+ \mathsfsl{D}+\eta_-(j) \xi_- \mathsfsl{E} +
  \eta_0(j) \mathsfsl{A}  \right) .
\end{equation}
Here we keep the restriction $N_++N_-<L$ for an admissible configuration and
$Z_L(\xi_+,\xi_-)$ is the normalization constant for this ensemble. Comparing with (\ref{MPA_C}),
it is obvious that the grand canonical measure is a superposition of the canonical measure
with coefficient proportional to $\xi_+^{N_+} \xi_-^{N_-}$.
In this grand canonical ensemble, the average density and the current
can be written in terms of the fugacities as
\begin{equation}
\begin{aligned}
\rho_{+,L}(\xi_+,\xi_-)
&= \frac{1}{L} \xi_+ \frac{\partial}{\partial \xi_+} \log Z_L(\xi_+,\xi_-),  \\
\rho_{-,L}(\xi_+,\xi_-)
&= \frac{1}{L} \xi_- \frac{\partial}{\partial \xi_-} \log Z_L(\xi_+,\xi_-) , \\
J_{+,L}(\xi_+,\xi_-)
&= \beta \langle \delta_{\eta_i,+} \delta_{\eta_{i+1},0}\rangle
    + \langle \delta_{\eta_i,+} \delta_{\eta_{i+1},-}\rangle
    -q \langle \delta_{\eta_i,-} \delta_{\eta_{i+1},+}\rangle \\
&= \left( \xi_+ \rho_{0,L-1} + \xi_- \rho_{+,L-1} + \xi_+ \rho_{-,L-1}   \right)
                   \times \frac{Z_{L-1}(\xi_+,\xi_-) }{Z_L(\xi_+,\xi_-) }, \\
J_{-,L}(\xi_+,\xi_-)
&= -\alpha \langle \delta_{\eta_i,0} \delta_{\eta_{i+1},-}\rangle
    - \langle \delta_{\eta_i,+} \delta_{\eta_{i+1},-}\rangle
    +q \langle \delta_{\eta_i,-} \delta_{\eta_{i+1},+}\rangle \\
&= -\left( \xi_- \rho_{0,L-1} + \xi_- \rho_{+,L-1}+ \xi_+ \rho_{-,L-1}  \right)
                     \times \frac{Z_{L-1}(\xi_+,\xi_-) }{Z_L(\xi_+,\xi_-)} .
\end{aligned}
\end{equation}
In the final formulas of the currents, obtained using  (\ref{DEA}),
it is understood that $\rho_{0,L} = 1-\rho_{+,L}-\rho_{-,L}$ and
$\rho_{\pm,L}$ should in fact mean $\rho_{\pm,L}(\xi_+,\xi_-)$.

The density and the current in the thermodynamic limit can be computed
by taking $L\to\infty$ of the above.
By a generalization of  the computations in~\cite{RSS00}, one finds, see Appendix~\ref{SectMatrixProduct},
\begin{equation}
\label{ZLnu}
 \lim_{L\to\infty}\frac{1}{L} \log Z_L(\xi_+,\xi_-) =\nu(\xi_+,\xi_-)
\end{equation}
with
\begin{equation}
 \nu(\xi_+,\xi_-) = \frac{(\xi_-+\sqrt{\xi_+ \xi_-} z)(\xi_++\sqrt{\xi_+ \xi_-} z)}{\sqrt{\xi_+ \xi_-} z},
  \label{nuxi}
\end{equation}
and
\begin{equation}
 z= z(\xi_+,\xi_-) = \frac{1+\xi_- a + \xi_+ b - \sqrt{(1+\xi_- a+\xi_+ b)^2-4ab\xi_+ \xi_-}}{ 2ab\sqrt{\xi_+ \xi_-}}
 \label{zxi}
\end{equation}
where  $a=-1+(1-q)/\alpha, b=-1+(1-q)/\beta$.
Then using these asymptotic expressions in the above, one finds
\begin{equation}\label{eq3.9}
\begin{aligned}
 \rho_{\pm}(\xi_+,\xi_-) &= \lim_{L\to\infty}  \rho_{\pm,L}(\xi_+,\xi_-)
 = \xi_{\pm} \frac{\partial}{\partial \xi_\pm} \log \nu(\xi_+,\xi_-),\\
 J_{\pm} (\xi_+,\xi_-) &= \lim_{L\to\infty}  J_{\pm,L}(\xi_+,\xi_-)
 = \pm (\xi_{\pm} -(\xi_{\pm}-\xi_{\mp})\rho_{\pm} )/\nu(\xi_+,\xi_-).
 \end{aligned}
\end{equation}

\textit{Remark:} For the special case $q=0$ the density-current relation is computed also in~\cite{Cantini} Eqs. (57), (58), avoiding intermediate chemical potentials.
Their equality with (\ref{eq3.9}) can be checked numerically (S. Prolhac, private communication).

The matrix $A$ defined in (\ref{15}) can be written in terms of the fugacities as
\begin{equation}\label{eqA}
A = \left(
      \begin{array}{cc}
\frac{\partial J_+}{\partial \xi_+} & \frac{\partial J_+}{\partial \xi_-} \\
\frac{\partial J_-}{\partial \xi_+} & \frac{\partial J_-}{\partial \xi_-}  \\
\end{array}
    \right)
    \left(
      \begin{array}{cc}
\frac{\partial \rho_+}{\partial \xi_+} & \frac{\partial \rho_+}{\partial \xi_-} \\
\frac{\partial \rho_-}{\partial \xi_+} & \frac{\partial \rho_-}{\partial \xi_-} \\
      \end{array}
    \right)^{-1}.
\end{equation}
One can also write the susceptibility matrix $C$ as derivatives of $\rho$,
\begin{equation}
 C =  \left(
      \begin{array}{cc}
         C_{++} & C_{+-} \\
         C_{-+}  & C_{--}   \\
\end{array}
    \right)
= \left(
      \begin{array}{cc}
         \xi_+ \frac{\partial \rho_+}{\partial \xi_+} & \xi_- \frac{\partial \rho_+}{\partial \xi_-} \\
         \xi_+ \frac{\partial \rho_-}{\partial \xi_+} &  \xi_- \frac{\partial \rho_-}{\partial \xi_-} \\
\end{array}
    \right)
\end{equation}
In terms of the fugacities, the condition  (\ref{15a}), $AC = CA^\mathrm{T}$, is equivalent to
\begin{equation}
 \xi_- \frac{\partial}{\partial \xi_-} J_-
 =
 \xi_+ \frac{\partial}{\partial \xi_+} J_+ .
\end{equation}
For the AHR model, this is easily  checked by noting that the currents can be written as
\begin{equation}
 J_{\pm} (\xi_+,\xi_-)
 =
\xi_{\pm}  \frac{\partial}{\partial \xi_\pm} \frac{\xi_+-\xi_-}{\nu(\xi_+,\xi_-)}.
\end{equation}

To switch to normal modes, we need to determine the matrix $R$ such that
\begin{equation}\label{eqRotated}
R A R^{-1} = \left(
               \begin{array}{cc}
                 v_1 & 0 \\
                 0 & v_2 \\
               \end{array}
             \right),
\quad
R C R^\mathrm{T} = \left(
            \begin{array}{cc}
              1 & 0 \\
              0 & 1 \\
            \end{array}
          \right).
\end{equation}
To obtain the matrix $R$, in general one has to compute the eigenvectors and eigenvalues of $A$.
The eigenvectors become the columns of the matrix called $(R^{0})^{-1}$.
For the AHR model, a simple expression of this matrix can be found by generalizating of the arguments in
\cite{KdN07} related to a quasi-particle picture.
It holds
\begin{equation}
\label{R0}
(R^{0})^{-1} =
\left(
  \begin{array}{cc}
  b \xi_+ - a b \sqrt{\xi_+ \xi_-} z(\xi_+,\xi_-) & b \xi_+ - \sqrt{\xi_+ \xi_-}/ z(\xi_+,\xi_-)  \\
  a \xi_- -  \sqrt{\xi_+ \xi_-} /z(\xi_+,\xi_-)        & a \xi_- - a b \sqrt{\xi_+ \xi_-} z(\xi_+,\xi_-) \\
  \end{array}
\right)
\end{equation}
and one verifies that
\begin{equation}
 R^{0} C (R^{0})^\mathrm{T} =  \left(
               \begin{array}{cc}
                 c_1 & 0 \\
                 0 & c_2 \\
               \end{array}
             \right)
\end{equation}
with $c_1,c_2>0$.
We still have the freedom of choosing the normalization of the eigenvectors. Defining the matrix $R$ as
$R_{\alpha\beta} = c_\beta^{1/2} (R^{(0)})_{\alpha\beta}$, $\alpha,\beta=1,2$, then Eq. (\ref{eqRotated}) is satisfied.

\section{Monte Carlo simulations of the AHR model}\label{SectMCSimulation}

\subsection{Exact sampling of the stationary initial condition}\label{SectIC}
For our Monte-Carlo simulations one has to numerically generate the stationary state. One standard method is to run the Monte-Carlo for ``sufficiently long'' time so to equilibrate towards the steady state. Because of the conservation laws, in our case ``long'' would mean diffusive time scale. Here we will use instead the method of exact sampling (\textit{i.e.}, a sampling without running the dynamics) exploiting the fact that the stationary measure of our system of size $L$ with periodic boundary conditions can be written using matrix product technique. This approach might be useful for other models with steady states of the same structure.

We specialize to the case $q=0$, $\alpha=\beta$, and equal fugacities $\xi_+=\xi_-=\xi$. Then (\ref{DEA}) becomes
\begin{equation}\label{eq1}
\mathsfsl{D} \mathsfsl{E} = \mathsfsl{D} + \mathsfsl{E},\quad \alpha \mathsfsl{D} \mathsfsl{A} = \mathsfsl{A}, \quad \alpha \mathsfsl{A} \mathsfsl{E} = \mathsfsl{A}.
\end{equation}
If we introduce $\mathsfsl{G}=\xi \mathsfsl{D} + \xi \mathsfsl{E} + \mathsfsl{A}$, the normalization $Z_L(\xi,\xi)$ is given by
\begin{equation}
Z_L(\xi,\xi) = \Tr'(\mathsfsl{G}^L)
\end{equation}
where, to meet the condition as in (\ref{eq3.4}) that $N_+ + N_- < L$, $\Tr'$ means the sum of the traces of all terms in the expansion of $G^L$ with at least one $\mathsfsl{A}$.

The connection between the matrices $\mathsfsl{A}$, $\mathsfsl{D}$,  $\mathsfsl{E}$ and the AHR $+,0,-$ particles is the following. The expansion of $\mathsfsl{G}^L$ is a linear combination of terms ordered products $\prod_{k=1}^L \mathsfsl{M_k}$ with $\mathsfsl{M_k}$ being one of the three matrices. The associated AHR particle configuration is such that:
\begin{equation}\label{eq4.3}
\begin{array}{ll}
\eta_+(k)=1,&\quad\textrm{ if }\mathsfsl{M_k}= \mathsfsl{D},\\
\eta_0(k)\,=1,&\quad\textrm{ if }\mathsfsl{M_k}= \mathsfsl{A},\\
\eta_-(k)=1,&\quad\textrm{ if }\mathsfsl{M_k}= \mathsfsl{E},
\end{array}
\end{equation}
and an AHR particle configuration has a weight given by $\xi^{(\#\mathsfsl{D}+\#\mathsfsl{E})}$.

As shown in~\cite{RSS00}, $\mathsfsl{A}$, $\mathsfsl{D}$, and $\mathsfsl{E}$ can be represented by semi-infinite matrices as follows: set $\gamma=\sqrt{1-(\alpha^{-1}-1)^2}$, then
\begin{equation}\label{eq4.4}
\mathsfsl{A}=\left(
           \begin{array}{cccc}
             1 & 0 & 0 & \cdots\\
             0 & 0 & 0 & \cdots\\
             0 & 0 & 0 & \cdots\\
             \vdots & \vdots & \vdots & \ddots\\
           \end{array}
         \right)
,\quad
\mathsfsl{D}=\left(
           \begin{array}{cccc}
             1/\alpha & \gamma & 0 & \cdots\\
             0 & 1 & 1 & \cdots\\
             0 & 0 & 1 & \cdots\\
             \vdots & \vdots & \vdots & \ddots\\
           \end{array}
         \right),
\quad
\mathsfsl{E}=\mathsfsl{D}^\mathrm{T}.
\end{equation}
These matrices can be interpreted as representing the transition weights (not yet transition probabilities!) of a random walk $X=(X_0,\ldots,X_L)$ on $\{0,1,2,\ldots\}$. Specifically, $\mathsfsl{A}$ means that we give weight $1$ if the walk stays at $0$ or zero otherwise, $\mathsfsl{D}$ gives a weight $\mathsfsl{D}_{k,k}$ if the walk stays at $k$, and weight $\mathsfsl{D}_{k,k+1}$ if the walk goes from $k$ to $k+1$, and similarly for $\mathsfsl{E}$.
The representation (\ref{eq4.4}) implies that
\begin{equation}\label{eq4.5}
\mathsfsl{G}=\left(
           \begin{array}{cccc}
             1+2 \xi/\alpha & \gamma \xi & 0 & \cdots\\
             \gamma \xi & 2\xi & \xi & \cdots\\
             0 & \xi & 2 \xi & \cdots\\
             \vdots & \vdots & \vdots & \ddots\\
           \end{array}
         \right).
\end{equation}

The above relation between random walks and matrices together with (\ref{eq4.3}) imply that $\mathsfsl{G}$ can be viewed as a matrix of the transition weights of a random walk $X=(X_0,\ldots,X_L)$ on $\{0,1,2,\ldots\}$, namely
\begin{enumerate}
  \item if $X_t=0$ and $X_{t+1}=0$, then at position $t+1$ we put a $+,0,-$ particle with probability proportional to $\xi/\alpha,1,\xi/\alpha$ respectively,
  \item if $X_t=k$ and $X_{t+1}=k$ for $k\geq 1$, then at position $t+1$ we put a $+,-$ particle with equal probability,
  \item if $X_t=k$ and $X_{t+1}=k+1$, then at position $t+1$ we put a $-$ particle,
  \item if $X_t=k$ and $X_{t+1}=k-1$, then at position $t+1$ we put a $+$ particle.
\end{enumerate}
Further, computing the trace corresponds to have a simple random walk which returns at the initial position after $L$ steps, \textit{i.e.}, with $X_0=X_L$.
Finally, since a $0$ particle occurs only if the random walk stays at position $0$ for one step, this means that if we want to impose the condition as in $\Tr'$, we need to have a random walk which has two consecutive $0$'s. By translation invariance we can first set a $0$ particle at position $t=1$ so that the requirement is satisfied. After the simulation we randomize the origin according to a uniform distribution on $\{1,\ldots,L\}$.

To simulate the random walk with weights $\mathsfsl{G}$, we first have to determine the corresponding stochastic matrix. This is obtained through the ground state transformation. First, let $E_0$ be the smallest (right) eigenvalue of $\mathsfsl{G}$ with eigenvector $\Omega=(\Omega_0,\Omega_1,\ldots)^{\mathrm{T}}$, \textit{i.e.}, satisfying
\begin{equation}
\mathsfsl{G}\, \Omega = E_0\, \Omega.
\end{equation}
Define the matrix $\mathsfsl{T}$ with entries
\begin{equation}
\mathsfsl{T}_{i,j}=\frac{1}{E_0 \Omega_i} \mathsfsl{G}_{ij} \Omega_j.
\end{equation}
Then, this is the stochastic matrix describing the random walk with transition weights given by entries of $\mathsfsl{G}$. As shown in~\cite{KdN07} $\Omega$ is given as follows:
\begin{equation}
E_0=\xi (\omega + 2+ 1/\omega),
\end{equation}
and
\begin{equation}
\Omega_0=1/\gamma,\quad \Omega_k=\omega^{k}, k\geq 1,
\end{equation}
where $\omega\in (0,1)$ is
\begin{equation}
\omega=\frac{2\xi}{1+2\xi (\alpha^{-1}-1)+\sqrt{1+4 \xi (\alpha^{-1}-1)}}.
\end{equation}

Having the stochastic matrix $\mathsfsl{T}$ is not quite enough for our purpose, since it gives us the transition probabilities of a random walk without the constraint that at time $L$ it has to return to $0$. This issue can be easily solved by the classical method of Doob-$h$ transform: let us define the function
\begin{equation}
h(x,t)=\Pb(X_L=0 \,| X_t=x) = (\mathsfsl{T}^{L-t})(x,0).
\end{equation}
Then, the Markov chain with law $\Pb_h$ and one-time transition probability given by
\begin{equation}\label{eqPh}
\Pb_h((x,t),(y,t+1))=\frac{1}{h(x,t)}\mathsfsl{T}(x,y) h(y,t+1)
\end{equation}
is exactly our random walk $X$ conditioned to be at $0$ at time $L$. To verify one multiplies (\ref{eqPh}) from time $t$ to time $L$,
\begin{equation}
\Pb_h\left(\bigcap_{s=t+1}^{L} \left\{X_s=x_s\right\}\bigg|X_t=x\right) = \frac{\mathsfsl{T}(x,x_{t+1})\cdots \mathsfsl{T}(x_{L-1},x_{L}) h(x_{L},L)}{h(x,t)},
\end{equation}
but $h(x_L,L)=\Pb(X_L=0\, | X_L=x_L)=\delta_{x_L,0}$, so that
\begin{equation}
\Pb_h\left(\bigcap_{s=t+1}^{L} \left\{X_s=x_s\right\}\bigg|X_t=x\right) = \frac{\mathsfsl{T}(x,x_{t+1})\cdots \mathsfsl{T}(x_{L-1},0)}{(\mathsfsl{T}^{L-t})(x,0)}\delta_{x_L,0}.
\end{equation}
Remark that to determine $h(x,t)$ we need only  $\mathsfsl{T}$ and $h(y,t+1)$, $y\geq 0$, since $h(x,t)=\sum_{y\geq 0} \mathsfsl{T}(x,y)h(y,t+1)$. For the simulation we do not use an analytic form for $\Pb_h$, rather we determine numerically the function $h$ and then $\Pb_h$.

To resume, in order to have an exact sampling of the initial condition, we have to do the following steps: (1) simulate a random walk with transition probability $\Pb_h$ given by (\ref{eqPh}) with $X_0=X_1=0$, (2) use the mapping described after (\ref{eq4.5}) to get an AHR particle configuration, (3) randomize the position of the origin.

\subsection{Normal modes and the theoretical scaling function}
In order to determine the normal modes and the theoretical prediction for the speed of the peaks and their width, we first need to determine the matrix $R$ such that
\begin{equation}
R A R^{-1} = \left(
               \begin{array}{cc}
                 v_1 & 0 \\
                 0 & v_2 \\
               \end{array}
             \right),
\quad
R C R^\mathrm{T} = \left(
            \begin{array}{cc}
              1 & 0 \\
              0 & 1 \\
            \end{array}
          \right).
\end{equation}
This can still be achieved analytically, as explained in Section~\ref{SectAHR}. $v_1$ and $v_2$ are the velocities of the peaks of the two normal modes. Now, to determine the spread we should compute the Hessian of $J_\pm$. Since $J_\pm$ are given as a function of the fugacities and not of the densities of particles, the computation is slightly involved. One follows (\ref{eqA}) but instead of one we have two derivatives. Define
\begin{equation}
A_2= \left(
    \begin{array}{cc}
      \frac{\partial \rho_+}{\partial\xi_+} & \frac{\partial \rho_+}{\partial\xi_-} \\
      \frac{\partial \rho_-}{\partial\xi_+} & \frac{\partial \rho_-}{\partial\xi_-} \\
    \end{array}
  \right),\quad J^{(1)}_\pm=A_2^{-1} \left(
                                  \begin{array}{c}
                                    \frac{\partial J_\pm}{\partial \xi_+} \\
                                    \frac{\partial J_\pm}{\partial \xi_-} \\
                                  \end{array}
                                \right).
\end{equation}
Then one can see that the Hessian matrices $H^+$ and $H^-$ are given by
\begin{equation}
H^\pm:= \left(
         \begin{array}{cc}
           \frac{\partial^2 J_\pm}{\partial \rho_+^2} & \frac{\partial^2 J_\pm}{\partial \rho_+\partial\rho_-} \\
           \frac{\partial^2 J_\pm}{\partial \rho_-\partial\rho_+} & \frac{\partial^2 J_\pm}{\partial \rho_-^2} \\
         \end{array}
       \right)
       = \left(
         \begin{array}{cc}
           \frac{\partial J^{(1)}_\pm}{\partial\xi_+} & \frac{\partial J^{(1)}_\pm}{\partial\xi_-} \\
         \end{array}
       \right) (A_2^{-1})^\mathrm{T}.
\end{equation}
According with (\ref{27}) the coupling matrices $G^1$ and $G^2$ are given by
\begin{equation}\label{eq4.18}
\begin{aligned}
G^1 &= \tfrac12 (R^{-1})^\mathrm{T} \left(R_{++} H^+ + R_{+-} H^-\right) R^{-1},\\
G^2 &= \tfrac12 (R^{-1})^\mathrm{T} \left(R_{-+} H^+ + R_{--} H^-\right) R^{-1},
\end{aligned}
\end{equation}
and the scaling coefficients are $\lambda_1=2\sqrt{2}|G^1_{11}|$ and $\lambda_2=2\sqrt{2}|G^2_{22}|$. All this can still be symbolically computed with Mathematica.

For the AHR model with $\xi_1=\xi_2=\xi$ and $\alpha=\beta$, we obtain the following expressions. Setting $a=\alpha^{-1}-1$, \mbox{$\eta=\sqrt{1+4\xi a}-1$}, the matrix $R$ is given by
\begin{equation}
R=\frac{-1}{2 (\eta +1)\sqrt{Q}}\left(
\begin{array}{cc}
\eta & \eta +2 \\
\eta +2 & \eta \\
\end{array}
\right),\quad Q=\frac{\eta  (\eta +2) (a +1) \left((\eta +2)^2 a -\eta ^2\right)}{4 (\eta +1)^3 ((\eta +2) a +\eta )^2},
\end{equation}
the velocity of the two peaks are
\begin{equation}
v_1=-v_2=\frac{4 (\eta +1) a  (\eta  (a -1)+2 a )}{(a +1) ((\eta +2) a +\eta ) \left((\eta +2)^2 a -\eta ^2\right)},
\end{equation}
and the spreading coefficients are
\begin{equation}
\lambda_1=\lambda_2=\left|\frac{4 \eta  (\eta +2) a  \left(\eta ^3-(\eta +4) \eta ^2 a +(\eta +2)^3 a ^3-(\eta -2) (\eta +2)^2 a ^2\right)}{\sqrt{Q/2}(a+1)\left(\eta ^3-(\eta +2)^3 a ^2-2 (\eta +2) \eta  a \right)^2}\right|.
\end{equation}

As can been seen from the coefficients $\lambda_i$, analytic formulas for the matrix elements of $G^1$ and $G^2$ tend to be lengthy. For purpose of comparison with other simulations, we report here the numerical values of these matrices for the parameters of Figures~\ref{FigSmallTimes}--\ref{FigR15} below. For $\xi=0.5$ and $\alpha=10/9$ we have
\begin{equation}\label{eq4.22}
G^1=\left(
\begin{array}{cc}
 -0.519 & -0.0369 \\
 -0.0369 & 0 \\
\end{array}
\right),\quad v_1=0.305,\quad \lambda_1=1.47,
\end{equation}
while for $\xi=0.5$ and $\alpha=2/3$ we have
\begin{equation}\label{eq4.23}
G^1=\left(
\begin{array}{cc}
 -0.341 & 0.113 \\
 0.113 & 0 \\
\end{array}
\right),\quad v_1=0.336,\quad \lambda_1=0.965,
\end{equation}
where $(G^2)_{\alpha\beta}=-(G^1)_{\beta\alpha}$, $v_1=-v_2$, and $\lambda_1=\lambda_2$.

\subsection{Numerical results}
Now we have all the ingredients for our theoretical prediction of the fitting functions. As before, let $S_{\alpha\beta}(j,t)=\langle\eta_\alpha(j,t) \eta_\beta(0,0)\rangle_{\vec{\rho}}-\rho_\alpha\rho_\beta$, $\alpha,\beta\in \{+,-\}$. Then, according to (\ref{28}), the correlation matrix of the two normal modes are given by
\begin{equation}
S^{\sharp}= R S R^{\mathrm{T}}.
\end{equation}
In $S^{\sharp}_{\alpha\alpha}$ there is a peak moving with velocity $v_\alpha$ and spread $(\lambda_\alpha t)^{2/3}$, for $\alpha=1,2$. Let $f_{\rm KPZ}$ be the KPZ function, see Figure~\ref{FigfKPZ}. The theoretical prediction for the fitting function of the peak in $S^{\sharp}_{\alpha\alpha}$ is given by
\begin{equation}\label{eq4.25}
(\lambda_\alpha t)^{-2/3} f_{\rm KPZ}\left((\lambda_\alpha t)^{-2/3}(x-v_\alpha t)\right).
\end{equation}

\begin{figure}
\begin{center}
\includegraphics[height=8cm]{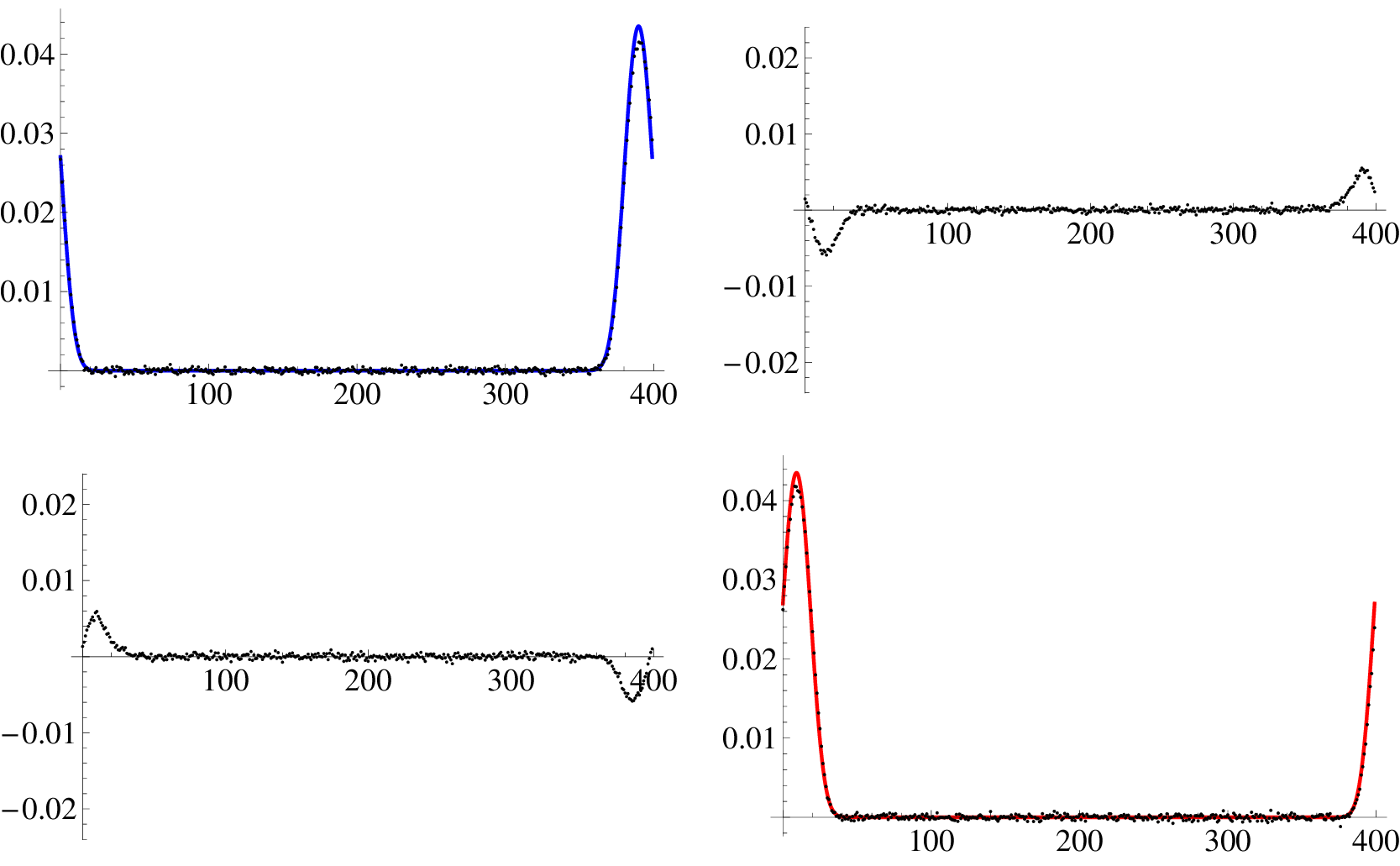}
\caption{Plot of the matrix elements $S^{\sharp}$ for small time, $t=30$, parameters $\xi=0.5$, $\alpha=10/9$, and $20\times 10^6$ MC runs. There is a signal in $S^{\sharp}_{12}$ and $S^{\sharp}_{21}$ distinct from the background noise, in the region where $S^{\sharp}_{11}$ and $S^{\sharp}_{22}$ still overlap.}
\label{FigSmallTimes}
\end{center}
\end{figure}
For the simulation we generate the initial condition using the exact sampling method explained in Section~\ref{SectIC}. This method allows us to have a fast sampling of the initial state without the uncontrolled errors resulting from equilibrating towards the steady state. In Figure~\ref{FigSmallTimes} we plot the  matrix elements of $S^{\sharp}$ for $t=30$, a short time when the two peaks have not yet separated. One sees that the off-diagonal elements still have a signal that is clearly distinguishable from the background noise. It is a dynamical effect and cannot be accounted for by the static correlations.

\begin{figure}[t]
\begin{center}
  \subfigure[$t=30$]{\includegraphics[width=0.45\textwidth]{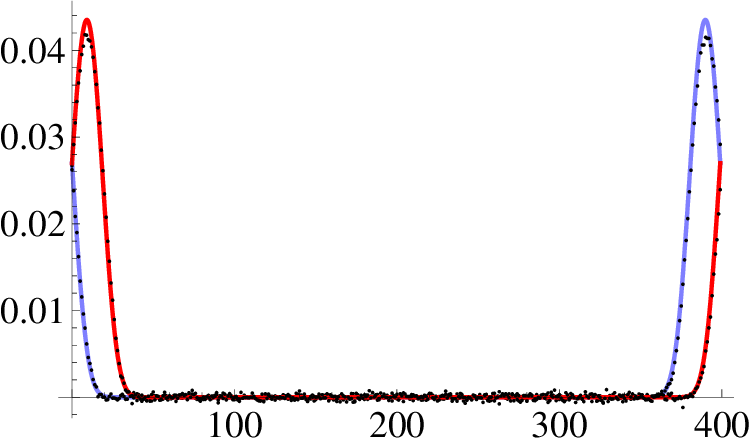}}\quad
  \subfigure[$t=150$]{\includegraphics[width=0.45\textwidth]{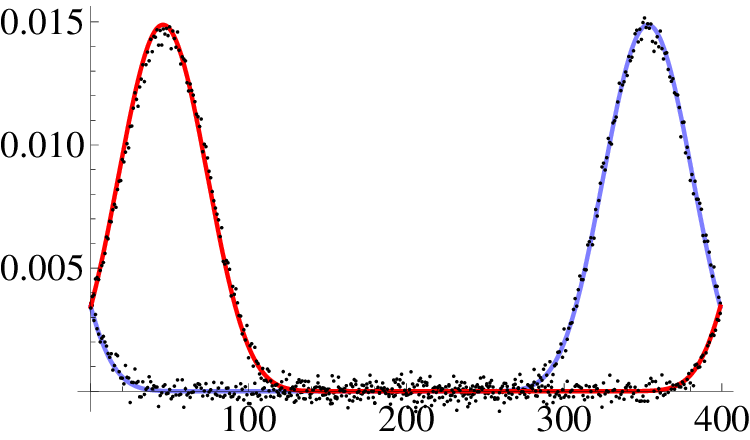}}\\
  \subfigure[$t=300$]{\includegraphics[width=0.45\textwidth]{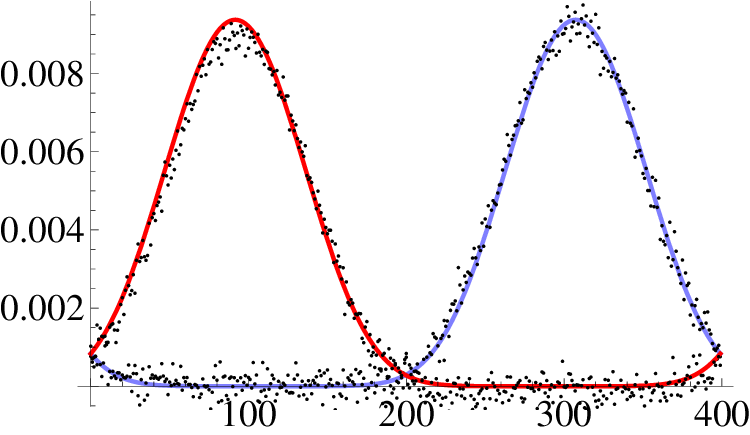}}\quad
  \subfigure[$t=450$]{\includegraphics[width=0.45\textwidth]{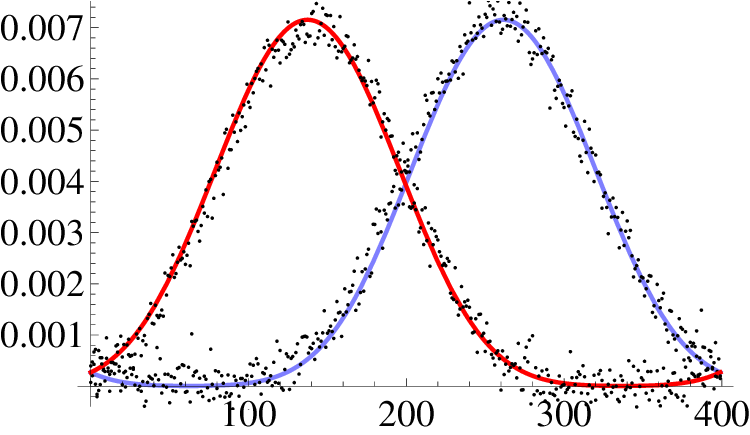}}
\caption{Plot of the matrix elements $S^{\sharp}_{11}$ (blue) and $S^{\sharp}_{22}$ (red) for different times and parameters $\xi=1.5$, $\alpha=10/9$, with an average over $20\times 10^6$ MC runs.}
\label{FigR09}
\end{center}
\end{figure}

\begin{figure}[t]
\begin{center}
  \subfigure[$t=30$]{\includegraphics[width=0.45\textwidth]{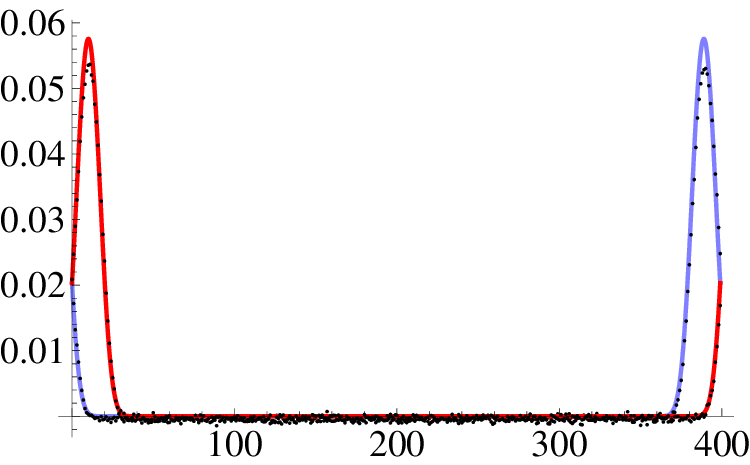}}\quad
  \subfigure[$t=150$]{\includegraphics[width=0.45\textwidth]{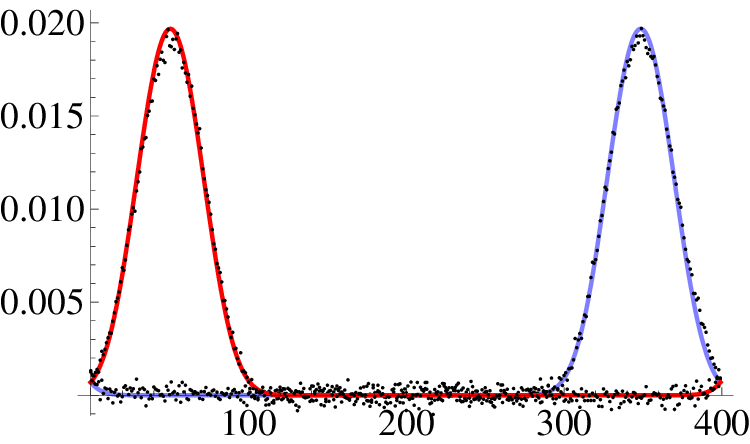}}\\
  \subfigure[$t=300$]{\includegraphics[width=0.45\textwidth]{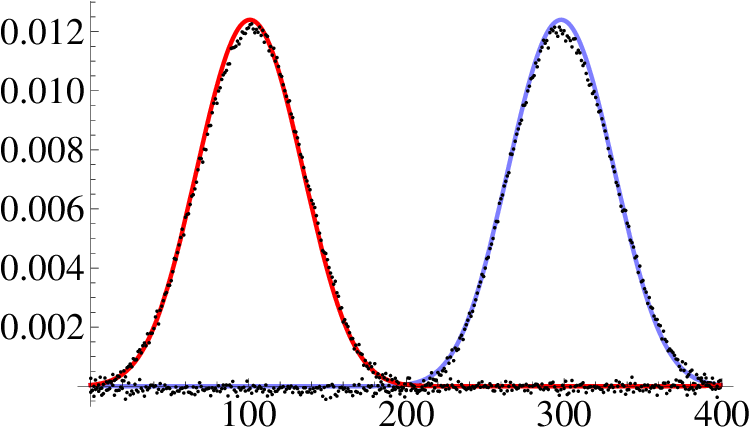}}\quad
  \subfigure[$t=450$]{\includegraphics[width=0.45\textwidth]{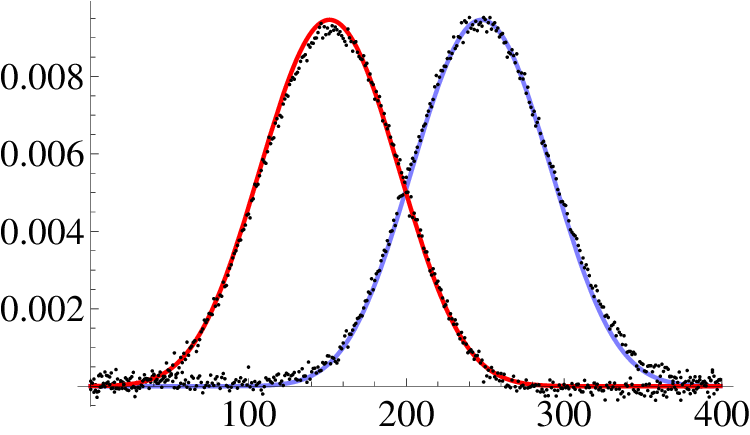}}
\caption{Plot of the matrix elements $S^{\sharp}_{11}$ (blue) and $S^{\sharp}_{22}$ (red) for different times and parameters
$\xi=1.5$, $\alpha=2/3$, with an average over $20\times 10^6$ MC runs for (a),(b), and $100\times 10^6$ MC runs for (c),(d).}
\label{FigR15}
\end{center}
\end{figure}

In Figures~\ref{FigR09} and~\ref{FigR15} we plot only the diagonal matrix elements of $S^{\sharp}$, in the same graph, for $4$ different times. For a single sample, one cannot discriminate the signal from noise. The signal appears only after an ensemble average over many Monte-Carlo runs. To have a good test of the theory the noise level has to be small. We average over $20\times 10^6$ MC runs. The theoretically predicted fitting function of the peaks is a very good approximation of the real data. Notice that there are no adjustable parameters. The fit  is to the theoretically predicted curves of  Eq.
(\ref{eq4.25}).

We have chosen the system size $L=400$ since it allows the two modes to separate before their peaks are in collision, which is the time span for which our theory applies. Surprisingly, the theoretical prediction is accurate even for longer time as one can see from Figures~\ref{FigR09}(d) and~\ref{FigR15}(d).
Eventually system size and peak width are the same and the fit (\ref{eq4.25}) stops being accurate.

Concerning the choice of the parameters, we have tested several densities and jump rates. Between them also the disorder point, $\alpha=\beta=1/2$, where the stationary measure is product measure. Qualitatively the results of the simulations do not change very much. In Figures~\ref{FigR09} and~\ref{FigR15} we have chosen equal density, since this makes the simulation faster. Also we have set \mbox{$\alpha=\beta\neq 1/2$} so that the initial steady state has non-trivial correlations. We have chosen two values of $\alpha$ such that the correlations in the initial condition are quite different. For instance, the correlation length at $\alpha=10/9$ is roughly twice as large as the one at $\alpha=2/3$.

We have also used the function (\ref{eq4.25}) with regarding $\lambda_1$ as free fit parameter. The agreement between the so obtained parameter $\lambda_1$ and the theoretically predicted one, (\ref{eq4.22}) for $\alpha=10/9$ (resp.~(\ref{eq4.23}) for $\alpha=2/3$), is fairly good already for relatively short times. At $t=30$ they differ already only by 5\% (resp.~7\%), with their difference slowly decreasing so that already at time $t=100$ it is below 3\% (resp.~5\%). In the one-component case, for related observables the slow decay is either as $\Or(t^{-1/3})$ or as $\Or(t^{-2/3})$~\cite{Tak12,TS12,FF11}.

The simulations were performed on a computer \emph{Intel(R) Xeon(R) CPU X5550 @ 2.67GHz}, which has $8$ processors. For the shorter simulation, \mbox{$t=30$}, $10^6$ Monte-Carlo runs were completed (by a single processor) after about $1$ hour, while for the longer simulation, $t=450$, one of the processor performed $10^6$ MC runs in about $11$ hours.

\section{Corrections to scaling}\label{SectCorrectionScaling}
To understand the speed of convergence to $f_{\mathrm{KPZ}}$, one has to analyze the subleading corrections to
(\ref{28}) on the basis of Eq.~(\ref{26}). Currently this can be achieved only on the level of the one-loop approximation to the stochastic field theory (\ref{26}),
which yields a closed equation for the correlator. Details of the argument can be found in~\cite{Spo13}, Appendix~B.
Let us define
\begin{equation}\label{29}
S^{\sharp\phi}_{\alpha\beta}(x,t) = \langle \phi_\alpha (x,t) \phi_\beta(0,0)\rangle.
\end{equation}
In one-loop approximation $S^{\sharp\phi}$ then satisfies the mode-coupling equation
\begin{equation}\label{B.20}
\begin{aligned}
\partial_t S^{\sharp\phi}_{\alpha\beta}(x,t)=& \sum^n_{\alpha'=1} \big(-v_\alpha\delta_{\alpha\alpha'}\partial_x +D_{\alpha\alpha'}\partial^2_x\big) S^{\sharp\phi}_{\alpha'\beta}(x,t)\\
&+ \int^t_0 ds \int_{\mathbb{R}} dy   \partial^2_y M_{\alpha\alpha'}(y,s) S^{\sharp\phi}_{\alpha'\beta}(x-y,t-s)
\end{aligned}
\end{equation}
with the memory kernel
\begin{equation}\label{B.21}
M_{\alpha\alpha'}(x,t)= \sum^n_{\beta',\beta'',\gamma',\gamma''=1} 2G^\alpha_{\beta'\gamma'} G^{\alpha'}_{\beta''\gamma''} S^{\sharp\phi}_{\beta'\beta''}(x,t) S^{\sharp\phi}_{\gamma'\gamma''}(x,t).
\end{equation}
Equations (\ref{B.20}) and (\ref{B.21}) have to be solved with the initial conditions \mbox{$S^{\sharp\phi}_{\alpha\beta}(x,0) =
\delta_{\alpha\beta}\delta(x)$}.

From numerical solutions of the mode-coupling equations one infers that the off-diagonal matrix elements of
$S^{\sharp\phi}$, while non-zero for short times, decay quickly to 0 at later times. Thus it seems to be safe to use the diagonal approximation
\begin{equation}\label{30}
S^{\sharp\phi}_{\alpha\beta}(x,t) = \delta_{\alpha\beta}f_\alpha (x,t),
\end{equation}
which leads to
\begin{equation}\label{31}
\partial_t f_\alpha(x,t)= (-c_\alpha \partial_x+D_\alpha \partial^2_x) f_\alpha (x,t) + \int^t_0 ds \int_{\mathbb{R}} dy
  f_\alpha(x-y,t-s) \partial^2_y M_{\alpha\alpha}(y,s)
\end{equation}
with memory kernel
\begin{equation}\label{32}
M_{\alpha\alpha}(x,t)= \sum_{\beta,\gamma=1} ^{n}2(G^\alpha_{\beta\gamma})^2 f_\beta(x,t) f_\gamma(x,t).
\end{equation}
For the solution to (\ref{31})  it is expected that $f_\alpha$ is well localized close to $v_\alpha t$.

Assuming for simplicity of the discussion that the $v_\alpha$'s are all distinct,
in the sum (\ref{32})  the terms $f_\beta f_\gamma$, $\beta\neq\gamma$, travel with distinct velocity  and hence can be neglected. This leads to the  approximation of the
memory kernel as
\begin{equation}\label{33}
\begin{aligned}
M_{\alpha\alpha}(x,t)&\simeq 2 (G^\alpha_{\alpha\alpha})^2 f_\alpha(x,t) f_\alpha(x,t)
+ \sum_{\beta=1,\beta\neq\alpha} ^{n}2(G^\alpha_{\beta\beta})^2 f_\beta(x,t) f_\beta(x,t)\\
&= M^0_{\alpha\alpha}(x,t)+M^1_{\alpha\alpha}(x,t).
\end{aligned}
\end{equation}
Dropping $M^1$, one is back to the decoupling discussed before. But now we have a tool to judge its accuracy.
The case of relevance in our context is $G^\alpha_{\alpha\alpha} \neq 0$ for all $\alpha$. Then to leading order, for every mode,
the long time behavior is computed with $M^0$. Making the scaling ansatz
\begin{equation}
f_\alpha(x,t)= (\lambda_\alpha t)^{-2/3} f_{\mathrm{mc}}((\lambda_\alpha t)^{-2/3}(x -  v_\alpha t)),
 \end{equation}
 for $f_{\mathrm{mc}}$ one arrives at the fixed point equation
 \begin{equation}\label{45a}
\tfrac{2}{3} \hat{f}'_{\mathrm{mc}}(w) = - \pi^2 w \int_0^1 ds \hat{f}_{\mathrm{mc}}((1-s)^{2/3}w) \int_{\mathbb{R}} dq
\hat{f}_{\mathrm{mc}}(s^{2/3}(w-q)) \hat{f}_{\mathrm{mc}}(s^{2/3} q),
\end{equation}
where $w \geq 0$ and $\,\hat{}\,$ denotes Fourier transform, $\hat{f}(k)= \int dx f(x) \exp(-\mathrm{i}2\pi xk)$. Numerically it is known that $f_{\mathrm{mc}}$ differs
 from $f_{\mathrm{KPZ}}$ only by order $5\%$~\cite{MS13}. To find the subleading correction we insert
 $f_\beta = f_{\mathrm{mc}}$ in  $M^1_{\alpha\alpha}$.
Then the memory kernel $M^1_{\alpha\alpha}$ is drifting relative to $v_\alpha t$ but it has a slow decay as $t^{-2/3}$, which is the origin for having
slowly decaying subleading terms. For more details we refer to~\cite{Spo13}.

For the AHR model the coupling matrices $G^1$, $G^2$ are written in (\ref{eq4.18}). Computationally one finds that
$G_{22}^1 = 0 = G_{11}^2$ for either $\xi_1 = \xi_2$ or $\alpha = \beta$. To our own surprise,  numerically these matrix elements seem to vanish always.
Hence the subleading terms vanish, which is the explanation for the observed rapid convergence in the Monte Carlo simulations.

By intention we have left out one subtle point in Section~\ref{sublattice}. If in (\ref{26}) one sets \mbox{$G^\alpha = 0$}, then the process is Gaussian and the steady state is easily computed to be white noise with covariance \mbox{$\langle \phi_\alpha(x,t)  \phi_{\alpha'}(x',t)\rangle = \delta_{\alpha\alpha'}\delta(x-x')$}. If the quadratic nonlinearity is added, in general one should be prepared for the steady state to change. However iff
\begin{equation}\label{34}
 G^\alpha_{\beta\gamma} =  G^\beta_{\alpha\gamma},
 \end{equation}
 then the steady state remains unchanged, see~\cite{Spo13}, Appendix B. For lattice gases the condition (\ref{34}) will not be satisfied generically. In particular, it does not hold for the AHR model. Now observe that we can freely choose  $G^\alpha_{\alpha\alpha}$,
$G^\alpha_{\beta\beta}$ and adjust the irrelevant terms such that  (\ref{34})  holds. In this sense, for the large scale behavior condition (\ref{34}) is not really a restriction.

\section{Summary}\label{SectSummary}
We simulated the two-component AHR model, which has the special feature that in the normal mode representation
the subleading coefficients $G_{22}^1 = 0 = G_{11}^2$. On the other hand,  the leading couplings $G^1_{11}$,
$G^2_{22} $ do not vanish and the two modes are still interacting through the sub-subleading terms
$G_{12}^1$, $G_{12}^2$, which are of the same order of magnitude as the leading terms. With very high precision we confirm
that the two modes satisfy KPZ scaling, including the scaling function and the dependence on the non-universal coefficients. Our results strongly support
the conjecture that nonlinear fluctuating hydrodynamics properly models the large scale behavior of the AHR model.

Two-component systems  have a much richer structure than their one-component cousin. The classification is naturally achieved through the velocities $v_1$, $v_2$ and the coupling matrices $G^1$, $G^2$. Based on examples
\cite{DBBR01,PS12}, if instead of three for AHR one allows for four states per lattice sites the entire manifold of admissible parameters can be
explored. It would be illuminating to study the case $G_{22}^1\neq 0$, $G_{11}^2\neq 0$, in order to better understand how subleading terms change the overall picture. From the
viewpoint of one-dimensional fluids, the conjunction of a KPZ and non-KPZ peak would be of great interest, \textit{i.e.},
$G^1_{11}\neq 0$, but $G^2_{22} =0$. Also, the original stepping stone~\cite{EK92} is very special from the general point of
view, since $v_1 = v_2$. Presumably the scaling functions are non-KPZ, but their precise shape has yet to be explored.
\bigskip\\
\textbf{Acknowledgement}. We thank H. van Beijeren and G. Sch\"utz for most informative discussions and J. Krug for pointing out the early literature on coupled KPZ equations. P.L.~Ferrari is grateful for the hospitality at the TU-Munich, where part of the work was made. His work is supported by the German Research Foundation via the SFB 1060--B04 project.

\appendix

\section{Proof of identity (\ref{15a})}\label{AppB}
As claimed in Eq.~(\ref{15a}), the linearized Euler currents, $A$, and the susceptibility matrix, $C$, satisfy
\begin{equation}\label{a1}
AC = CA^\mathrm{T}.
\end{equation}
This relation is well known for classical fluids, see \textit{e.g.}~\cite{Spo91}, and for anharmonic chains~\cite{Spo13}.
For many-component lattice gases it was noted by T\'{o}th and Valk\'{o}~\cite{TV03} in a special case and proved in generality  by Grisi and Sch\"{u}tz~\cite{GSch12}. Here we give a simple proof which relies only on the conservation laws and on space-time stationarity.

The conserved fields are denoted by  $\eta_\alpha(j,t)$, $j\in\mathbb{Z}$, $t \in\mathbb{R}$, $\alpha = 1,\ldots,n$.
$\eta_\alpha(j,t)$ is a space-time stationary process with zero mean. By stationarity
\begin{equation}\label{App1}
S_{\alpha\beta}(j,t) = \langle  \eta_\alpha(j,t) \eta_\beta(0,0)\rangle = S_{\beta\alpha}(-j,-t).
\end{equation}
Using the conservation law,
\begin{equation}
\begin{aligned}
 \frac{d}{dt}\sum_{j \in \mathbb{Z}} jS_{\alpha\beta}(j,t) &=  \sum_{j \in \mathbb{Z}} j
\langle   (\mathcal{J}_\alpha(j-1,t) -   \mathcal{J}_\alpha(j,t))\eta_\beta(0,0)\rangle \\
&= \sum_{j \in \mathbb{Z}} \langle   \mathcal{J}_\alpha(j,t)\eta_\beta(0,0)\rangle =   \sum_{j \in \mathbb{Z}} \langle   \mathcal{J}_\alpha(0,0)\eta_\beta(-j,-t)\rangle \\
&= \sum_{j \in \mathbb{Z}} \langle   \mathcal{J}_\alpha(0,0)\eta_\beta(j,0)\rangle.
\end{aligned}
\end{equation}
As discussed in \cite{Ku84,GSch12}, the infinite volume average can be obtained from a system on a ring by introducing
a chemical potential, $\mu_\alpha$, for the density $\rho_\alpha$ in the conventional way.  Hence
\begin{eqnarray}\label{App2}
&&\hspace{0pt}\sum_{j \in \mathbb{Z}} \langle   \mathcal{J}_\alpha(0,0)\eta_\beta(j,0)\rangle =
\frac{\partial}{\partial \mu_{\beta}} \langle   \mathcal{J}_\alpha(0,0)\rangle_{\vec{\mu}}\\
&&\hspace{0pt}= \sum_{\gamma =1}^{n} \frac{\partial}{\partial \rho_{\gamma}} \langle   \mathcal{J}_\alpha(0,0)\rangle_{\vec{\rho}} \,\frac{\partial \rho_{\gamma}}{\partial \mu_{\beta}} = (AC)_{\alpha\beta}
\end{eqnarray}
and
\begin{equation}
\sum_{j \in \mathbb{Z}} jS_{\alpha\beta}(j,t) =  (AC)_{\alpha\beta}t +\sum_{j \in \mathbb{Z}} jS_{\alpha\beta}(j,0) .
\end{equation}
Summing in Eq.~(\ref{App1}) over $j$ yields
\begin{equation}
(AC)_{\alpha\beta}t = (AC)_{\beta\alpha}t,
\end{equation}
which is the desired identity.

\section{The asymptotics of $Z_L(\xi_+,\xi_-)$}\label{SectMatrixProduct}
To study the asymptotics of $Z_L(\xi_+,\xi_-)$, it is useful to introduce its generating function,
\begin{equation}
 \Theta(\lambda) := \sum_{L=0}^{\infty} \lambda^L Z_L(\xi_+,\xi_-).
\end{equation}
By a simple generalization of the arguments in~\cite{RSS00}, this can be written as
\begin{equation}
 \Theta(\lambda)
 =
 \frac{\frac{d}{d\lambda} \left( \lambda \Theta^{(0)}(\lambda)\right )} {1-\lambda \Theta^{(0)}(\lambda)}
\end{equation}
where
\begin{gather}
\mathsfsl{C} = \xi_+ \mathsfsl{D} + \xi_- \mathsfsl{E}, \\
 Z_L^{(0)}(\xi_+,\xi_-) = \langle 0|\mathsfsl{C}^L |0\rangle, \\
 \Theta^{(0)}(\lambda) =  \sum_{L=0}^{\infty} \lambda^L Z_L^{(0)}(\xi_+,\xi_-)
\end{gather}
with $|0\rangle = (1,0,0,\ldots )^\mathrm{T}$. The asymptotics of $Z_L(\xi_+,\xi_-)$ is understood by knowing
the solution of
\begin{equation}
 \lambda \Theta^{(0)}(\lambda) = 1
 \label{lT1}
\end{equation}
which is closest to the origin. If we denote it by $1/\nu(\xi_+,\xi_-)$, we conclude (\ref{ZLnu}).

For $Z_L^{(0)}(\xi_+,\xi_-)$, one can find an integral formula~\cite{Sas00},
\begin{equation}
Z_L^{(0)}(\xi_+,\xi_-)
 = \frac{(q,ab;q)_{\infty}}{4\pi \mathrm{i}} \int_{\mathcal{C}} \frac{dz}{z}
 \frac{(z^2,z^{-2};q)_{\infty}[(\sqrt{\xi_-}+\sqrt{\xi_+} z)(\sqrt{\xi_+}+\sqrt{\xi_-} /z)]^L}
        {(a\sqrt{\xi_-/\xi_+} z, a\sqrt{\xi_+/\xi_-} /z,b\sqrt{\xi_+/\xi_-} z, b\sqrt{\xi_-/\xi_+} /z;q)_{\infty}},
\end{equation}
where  we introduced the notations,
\begin{equation}
 (a;q)_{\infty} = \prod_{i=0}^{\infty} (1-a q^i), \quad
 (a_1,\ldots, a_k;q)_{\infty} = \prod_{i=1}^k (a_i;q)_{\infty},
\end{equation}
and the contour $\mathcal{C}$ of the integral should include the poles at \mbox{$z=0, a \sqrt{\xi_+/\xi_-}q^k,  b \sqrt{\xi_-/\xi_+}q^k$},
$k=0,1,2,\ldots$.
From this one gets an integral formula for $\Theta^{(0)}(\lambda)$:
\begin{equation} \label{ThetaInt}
\begin{aligned}
 \Theta^{(0)}(\lambda)
 &=
 \frac{(q,ab;q)_{\infty}}{4\pi \mathrm{i}} \int_{\tilde{\mathcal{C}}} \frac{dz}{z}
 \frac{(z^2,z^{-2};q)_{\infty}}
        {(a\sqrt{\xi_-/\xi_+} z, a\sqrt{\xi_+/\xi_-} /z,b\sqrt{\xi_+/\xi_-} z, b\sqrt{\xi_-/\xi_+} /z;q)_{\infty}}\\
 &\quad\times     \frac{1}{ (1- (\sqrt{\xi_-}+\sqrt{\xi_+} z)(\sqrt{\xi_+}+\sqrt{\xi_-} /z)/\lambda)},
\end{aligned}
\end{equation}
where the contour $\tilde{\mathcal{C}}$ now also includes the smaller pole coming from the last factor in (\ref{ThetaInt}).
When $q=0$, this contour integral can be evaluated explicitly with the result,
\begin{equation}
 \Theta^{(0)}(\lambda)
 =
 \frac{z(\xi_+,\xi_-)}{\lambda\sqrt{\xi_+\xi_-}(1-a\sqrt{\xi_-/\xi_+} z(\xi_+,\xi_-))(1-b\sqrt{\xi_+/\xi_-}z(\xi_+,\xi_-))},
\end{equation}
where $z(\xi_+,\xi_-)$ is given by (\ref{zxi}).
If we consider the solution to (\ref{lT1}) for this choice of $ \Theta^{(0)}(\lambda)$, one arrives (\ref{nuxi}) with (\ref{zxi}).

\end{document}